\pdfoutput=1
\documentclass[11pt,a4paper]{article}

\usepackage[margin=1in]{geometry}
\usepackage[T1]{fontenc}
\usepackage[utf8]{inputenc}
\usepackage{lmodern}
\usepackage[authoryear,round]{natbib}
\usepackage{graphicx}
\usepackage{booktabs}
\usepackage{tabularx}
\usepackage{longtable}
\usepackage{array}
\usepackage{setspace}
\usepackage{xcolor}
\usepackage{listings}
\usepackage{textcomp}
\usepackage{amsmath}
\usepackage{authblk}
\usepackage{caption}
\usepackage{url}
\usepackage{xurl}
\usepackage{float}
\usepackage[hidelinks]{hyperref}

\lstdefinestyle{lcsi}{
  basicstyle=\ttfamily\scriptsize,
  breaklines=true,
  breakatwhitespace=false,
  columns=fullflexible,
  frame=single,
  framesep=4pt,
  showstringspaces=false,
  keepspaces=true,
  showtabs=false,
  tabsize=2,
  morecomment=[l]{\#},
  commentstyle=\itshape\color{gray},
  xleftmargin=2pt,
  xrightmargin=2pt,
  extendedchars=true,
  literate=
    {′}{{$^\prime$}}{1}
    {ü}{{\"u}}{1}
    {γ}{{$\gamma$}}{1}
    {θ}{{$\theta$}}{1}
    {α}{{$\alpha$}}{1}
    {…}{{...}}{3}
    {—}{{--}}{2}
    {–}{{--}}{2}
    {→}{{$\rightarrow$}}{1}
}
\title{LitCurate: A Configuration-Driven AI-Assisted Framework for Scientific Database Construction with an Application to Lower-Mantle Equation-of-State Data}

\author[1]{Abin Shakya\thanks{Corresponding author: ashakya@ldeo.columbia.edu}}
\author[2]{Wilson Samuels}
\author[2]{Dominica Wilson}
\author[3]{Gioia A. Marchi}
\author[4]{Israa Draz}
\author[4]{Chenxing Luo}
\author[1,4,5]{Renata M. Wentzcovitch\thanks{Corresponding author: rmw2150@columbia.edu}}

\affil[1]{Lamont--Doherty Earth Observatory, Columbia University, Palisades, NY, 10964, United States}
\affil[2]{Department of Electrical and Computer Engineering, Tuskegee University, Tuskegee, AL, 36088, United States}
\affil[3]{Department of Industrial Engineering and Operations Research, Columbia University, New York, NY, 10027, United States}
\affil[4]{Department of Applied Physics and Applied Mathematics, Columbia University, New York, NY, 10027, United States}
\affil[5]{Department of Earth and Environmental Sciences, Columbia University, New York, NY, 10027, United States}

\date{}

\begin{document}
\maketitle

\begin{abstract}
The growing scientific literature contains decades of experimental and computational results that could support data-driven and physics-based modeling, yet much of this information remains locked in publications and is not readily usable for large-scale analysis or scientific software. Building structured databases from the literature is particularly challenging when relevant studies must first be discovered among large collections of papers and reported quantities must be extracted with enough scientific context to remain usable. We present LitCurate, an open-source framework for building scientific databases from the literature using large language models within an auditable, stage-wise curation workflow. LitCurate integrates literature discovery, relevance screening, full-text processing, and structured information extraction while retaining intermediate results and provenance, allowing researchers to inspect and revise individual stages rather than treating automated curation as a black-box process. We apply LitCurate to construct an equation-of-state database of lower-mantle and lower-mantle-relevant high-pressure mineral phases from experimental and theoretical studies, comprising 1{,}334 entries from 205 papers. The resulting dataset links reported equation-of-state parameters to mineral phases, compositions, equation formulations, methods, and parameter constraints, and labels values as source-reported or citation-reported when provenance can be determined. The records are available through a searchable web application. By connecting scientific literature to traceable, machine-readable data, LitCurate provides a reusable approach for transforming accumulated literature into resources for scientific analysis and computational modeling.
\end{abstract}


\section{Introduction}
\label{sec:introduction}
The scientific literature encodes decades of experimental measurements, theoretical calculations, and interpretive models that could support both data-driven discovery and large-scale analysis. Initiatives such as Deep-time Digital Earth and broader reviews of Earth artificial intelligence emphasize that progress increasingly depends on open, interoperable, machine-readable data beyond publications alone \citep{wang2021deep,sun2022earth}. Yet across disciplines, much of the primary evidence remains locked in journal articles, where it is distributed across narrative text and tables, and is not available as structured databases with retained provenance. Building such databases from the literature is therefore a recurring bottleneck. Relevant studies must first be identified within large, heterogeneous collections of papers, and reported quantities must be interpreted within a scientific context before they can be standardized for reuse. Without explicit provenance and intermediate artifacts, automated curation risks becoming a black-box process that is difficult to audit, revise, or reproduce.

These challenges are not unique to any single field. In this work, we address them in the context of mineral physics by focusing on equation-of-state (EOS) parameters for lower-mantle minerals. These parameters characterize compression behavior and constrain density and bulk modulus at high pressures; when combined with complementary thermal and shear-elastic properties, they contribute to mineral-physics models used to interpret mantle composition, thermal state, and seismic observations \citep{mattern2005lower,murakami2012perovskitic,wentzcovitch2004thermoelastic,wentzcovitch2010thermodynamic,wentzcovitch2010thermoelasticity}. Key EOS quantities such as the zero-pressure volume ($V_0$), isothermal bulk modulus ($K_0$), and its pressure derivative ($K_0'$) are reported across experimental compression studies and first-principles calculations using diverse formulations such as Birch--Murnaghan \citep{murnaghan1944compressibility,birch1947finite} and Vinet \citep{vinet1987compressibility}, together with heterogeneous conventions for units, reference conditions, and fixed versus fitted parameters. Thermodynamic frameworks that synthesize mineral properties at mantle conditions \citep{stixrude2005thermodynamics,stixrude2011thermodynamics,stixrude2024thermodynamics,connolly2009geodynamic}, first-principles calculations of high-pressure elasticity for the major mantle phases \citep{karki2001elastic}, and open computational toolkits such as BurnMan \citep{cottaar2014burnman,myhill2023burnman} and MAGEMin \citep{riel2022magemin} demonstrate how standardized mineral parameters become executable inputs for mantle-property calculations. In practice, however, assembling those parameters from primary literature remains labor-intensive, difficult to maintain as new studies appear, and hard to reproduce when extraction decisions are not recorded. The scientific need is therefore not only for extracted numbers, but for records that retain phase, composition, method, and constraint metadata alongside literature provenance so that reported values can be compared across studies and downstream models can be updated transparently.

Prior work has developed platforms to accelerate the construction of literature-derived databases. GeoDeepDive combined document aggregation with high-throughput text parsing to support machine reading over large literature corpora \citep{peters2017geodeepdive}. More recent systems emphasize human--AI collaboration for multimodal extraction from PDFs: GeoDeepShovel supports annotation and extraction from tables, figures, and text for Deep-Time Digital Earth database projects \citep{zhang2023geodeepshovel}, and GeoKnowledgeFusion targets joint understanding of text, images, and tables for compiling domain databases such as isotope records \citep{guo2024geoknowledgefusion}. These platforms substantially reduce the cost of multimodal extraction once a document collection is available, and they are particularly effective when experts interactively annotate known PDFs. They less often provide a complete, reusable package for discovery-to-export curation in which users choose LLM providers and models, configure prompts and schemas, and retain stage-wise intermediates for inspection and resumption.

Large language models (LLMs) have emerged as flexible tools for structured information extraction from scientific literature \citep{dagdelen2024structured}. Domain-specific efforts have demonstrated their potential to populate substantial property databases. In polymer science, \citet{paudel2026polymer} screened 48{,}206 abstracts to identify 198 property-reporting publications and extracted 1{,}095 records of properties obtained from molecular-dynamics simulations by combining LLM-assisted extraction with expert curation. Such efforts, however, are generally designed around individual research questions, with search terms, screening logic, and extraction targets tailored to a particular study. Complementing these domain-specific applications, more general scientific extraction systems have emphasized schema flexibility and modular extraction workflows. SciDaSynth uses LLMs to generate schema-shaped tables from user queries and supports interactive validation of inconsistencies across documents \citep{wang2024scidasynth}. SciEx decouples PDF parsing, multimodal retrieval, schema-guided extraction, and aggregation while allowing models and prompting strategies to be exchanged for on-demand extraction \citep{li2025sciex}. Although both systems advance schema flexibility and support adaptable extraction workflows, they typically assume a preassembled document collection and emphasize interactive or modular extraction and aggregation. A further limitation shared by these systems concerns multimodal content. Text and tables are increasingly tractable, whereas scientific figures remain challenging. Current systems often require substantial manual annotation for figure extraction, and automated information recovery across diverse plot types has not yet achieved the reliability required for scientific database construction \citep{he2026exchart, li2025sciex, davila2021chart}.

What remains comparatively underdeveloped is an open-source end-to-end pipeline that combines literature discovery with schema-guided LLM extraction while allowing domain knowledge to be specified through configurable prompts and schemas instead of being embedded in the pipeline code. Open catalogs such as OpenAlex \citep{priem2022openalex} make programmatic discovery feasible at scale, but discovery alone does not yield curated, structured data. Conversely, LLM extraction alone does not solve document selection, PDF access gaps, heterogeneous reporting, or the operational needs of long-running curation projects. Bridging these capabilities requires an inspectable sequence of stages, so that search, screening, and extraction can be examined and re-run separately.

Here we present LitCurate, an open-source package for building scientific databases from the literature using LLMs within an auditable, stage-wise workflow. A run proceeds from literature search through relevance screening and full-text preparation to schema-guided extraction and export. Intermediate artifacts and a run ledger record each stage so that work can be inspected, re-run, or resumed. Domain knowledge (research goals, screening criteria, extraction schemas, and prompts) lives in configuration files, so the codebase can be adapted across disciplines without rewriting pipeline stages. Extraction operates on converted text and tables.

We demonstrate LitCurate by compiling a database of EOS parameters for lower-mantle and lower-mantle-relevant high-pressure mineral phases from experimental and theoretical studies, comprising 1{,}334 entries from 205 papers. Each record links reported parameters to phase, composition, formulation, method, and fitting constraints, and, where provenance can be determined, distinguishes new determinations from values that a paper quotes from earlier studies. The records can be explored through a searchable web application (\url{https://eos.litcurate.com}).

\section{Methodology}
\label{sec:methodology}
\begin{figure}
    \centering
    \includegraphics[
        width=\textwidth,
        height=1.0\textheight,
        keepaspectratio
    ]{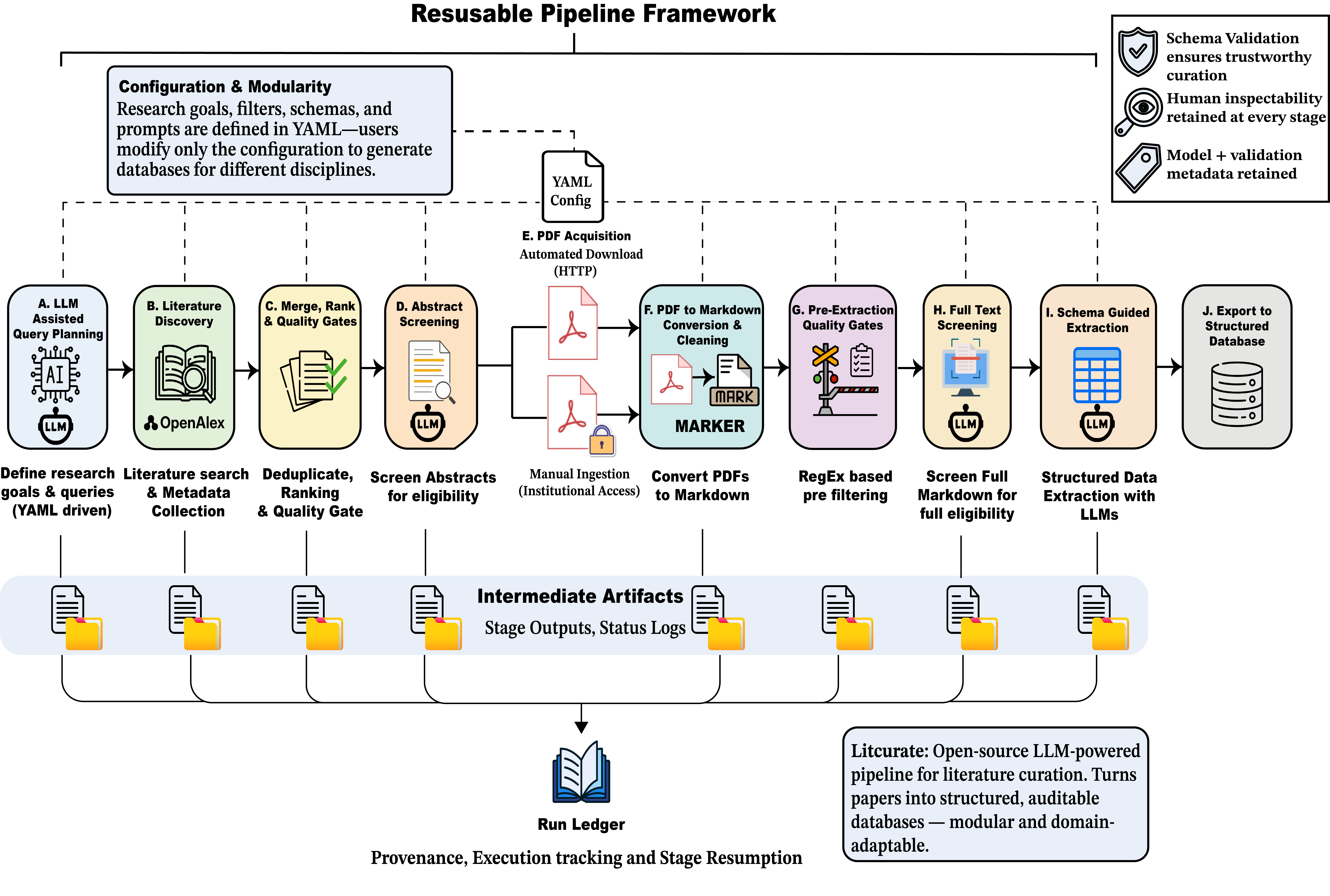}
    \caption{LitCurate architecture. Pipeline advances through ten stages (A--J), from LLM-assisted query planning and literature discovery through screening, full-text preparation, and schema-guided extraction, to export of a merged structured database. Each stage writes intermediate artifacts and records its status in a per-run ledger (bottom layer), so individual stages can be inspected, re-run, or resumed. Domain knowledge (research goals, filter criteria, extraction schemas, and prompts) is supplied via YAML configuration and associated files (top-left), allowing stage machinery to be reused across disciplines.}
    \label{fig:architecture}
\end{figure}

\subsection{Architectural overview}
\label{sec:architecture}

LitCurate is an open-source Python package that decomposes literature-to-database curation into a fixed sequence of stages (Fig.~\ref{fig:architecture}A--J). Each stage reads the artifacts produced by earlier stages, writes new artifacts under a per-run directory, and records its status in a SQLite \citep{hipp2020sqlite} run ledger. Stages can be executed one at a time or as a full pipeline from a single command, and a run that fails or is interrupted can be resumed from the last completed stage. Each run directory also stores a frozen copy of the configuration used, so a completed run remains interpretable after the working configuration has changed; a revised criterion or prompt is applied by resetting that stage and re-running it and its downstream stages. Domain knowledge (research goal, screening criteria, extraction schemas, and prompts) is defined in YAML and associated files, separate from the pipeline code. Model choice is per-stage, so commercial APIs and self-hosted open-weight models can be combined in one run. Every extraction is validated against its declared schema and stored with the model identity, timestamp, and validation verdict that produced it.

\subsection{Pipeline stages}
\label{sec:stages}

A run begins with a natural-language research goal specified in the configuration (Fig.~\ref{fig:architecture}A). During query planning, the configured LLM translates this broad goal into a set of focused search queries that cover different target materials, properties, methods, and terminology. The generated queries are saved as an editable artifact, allowing users to review or modify them before retrieval begins. LitCurate submits each query independently to OpenAlex across the configured publication-year intervals and preserves the returned records for every query (Fig.~\ref{fig:architecture}B). Because a paper may be retrieved by multiple queries, the merge-and-rank stage combines all results, removes duplicate papers, applies configurable quality requirements—such as a DOI, an abstract, and an eligible publication type—and ranks the remaining candidates by search frequency, relevance, and citation information. The result is a single-ranked table of candidate papers and their bibliographic metadata (Fig.~\ref{fig:architecture}C). For a detailed illustration of this discovery workflow, see Fig.~\ref{fig:s_discovery}.

Candidates are next screened at the abstract level against the domain-specific criteria defined in the configuration (Fig.~\ref{fig:architecture}D). The LLM assigned to this stage records a keep-or-exclude decision for each paper, together with its justification, allowing PDF acquisition and conversion to focus on studies likely to contain extractable data. For papers retained by this screening, the acquisition stage (Fig.~\ref{fig:architecture}E) uses metadata from Unpaywall \citep{piwowar2018state} and OpenAlex \citep{priem2022openalex} to identify candidate full-text locations and attempts to download the PDFs through direct HTTP requests. When automated retrieval does not return a file, LitCurate also allows users to add PDFs obtained through institutional access in accordance with publisher access policies. Automatically retrieved and manually added PDFs are registered in one download manifest and follow that downstream workflow. The manifest records the acquisition outcome for every paper, including whether the PDF was obtained, unavailable, or skipped. In the present application, PDFs were obtained for 918 of the 1{,}030 papers retained after abstract screening.

Acquired PDFs are converted to markdown with Marker \citep{paruchuri2024marker}, a document-conversion tool that reconstructs reading order and renders tables as text (Fig.~\ref{fig:architecture}F). The conversion device is configurable (CPU or CUDA); GPU execution is recommended for large corpora. A cleaning step then strips configured back matter (references, acknowledgments, funding statements, and optionally supplementary sections) to produce extraction-ready text. Both the raw and cleaned markdown are retained, and the configuration determines which of the two is supplied to extraction, so the effect of cleaning is itself inspectable. Two gates sit between text preparation and extraction: a pre-extraction gate (Fig.~\ref{fig:architecture}G) applies inexpensive regular-expression signals to skip papers whose full text shows no evidence of the target quantities, avoiding LLM cost on documents that cannot yield records, and papers that pass are screened once more at the full-text level (Fig.~\ref{fig:architecture}H) against the configured criteria, now with access to the complete converted document instead of only the abstract. A stage-level view of screening, acquisition, conversion, and these two gates is given in Fig.~\ref{fig:s_screening}.

\begin{figure}
    \centering
    \includegraphics[
        width=0.9\textwidth,
        height=0.9\textheight,
        keepaspectratio
    ]{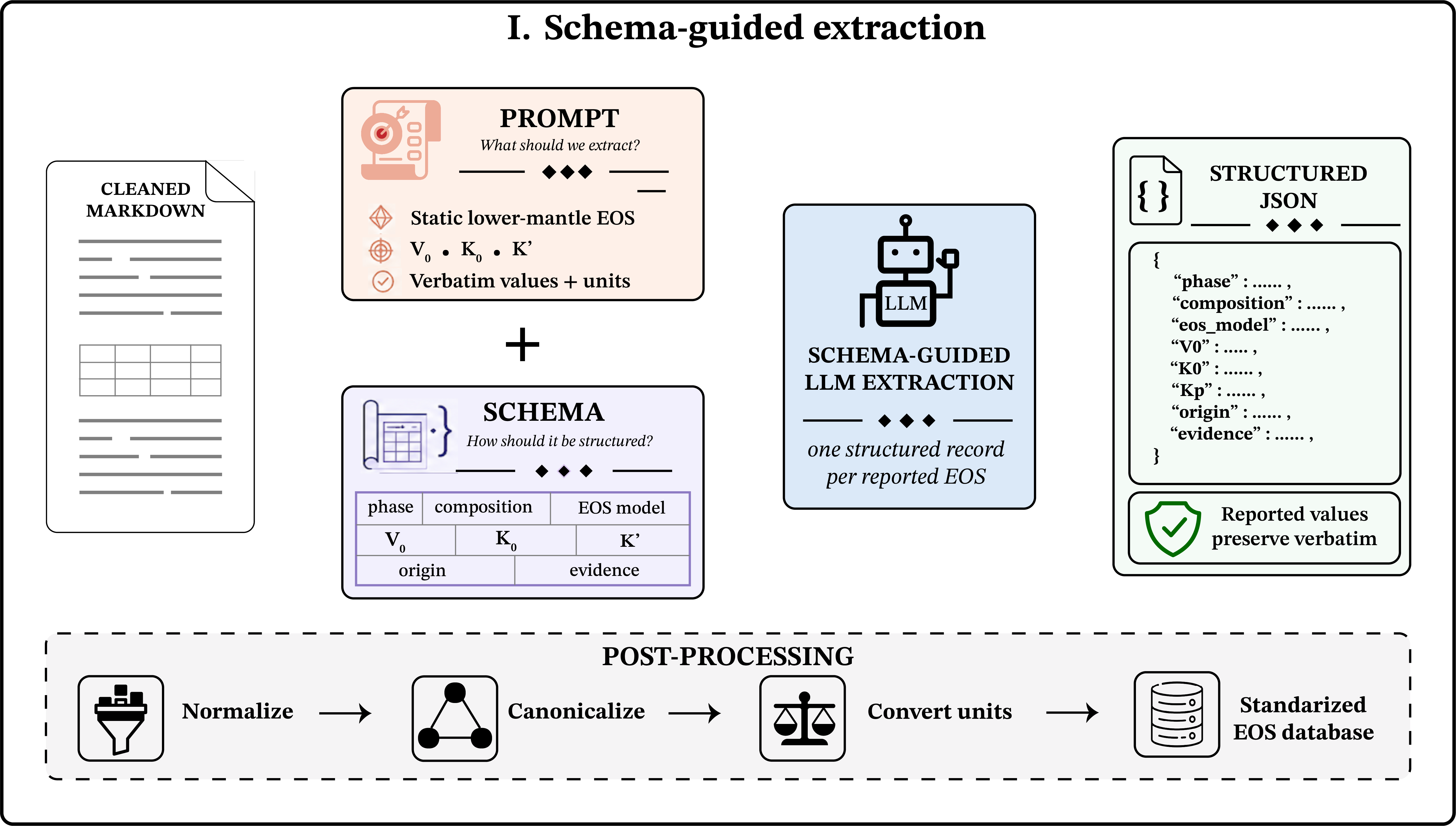}
    \caption{Schema-guided extraction (stage I). Cleaned markdown is combined with a task prompt and a declared record schema; LLM returns structured JSON that is validated against that schema and stored as an envelope, then merged into exported EOS database.}
    \label{fig:extraction}
\end{figure}

Extraction (Fig.~\ref{fig:extraction}) is driven by schema--prompt pairs declared in the configuration. A schema defines the shape of the records to be produced (fields, types, and required properties) and may be authored in JSON Schema or in a compact, declarative YAML form; the paired prompt instructs the model on how to populate that shape from the paper. The lower-mantle schema also includes an open extra-information object for reported quantities that are not first-class fields, including the pressure scale or calibrant when the paper reports one. For each eligible paper, the configured LLM receives the prepared text and the extraction instructions, and must return a JSON response that LitCurate validates against the declared schema. The result is stored as an \emph{envelope}: the payload together with the schema name and version, the model identifier, a timestamp, and the validation verdict with any errors.

Finally, the export stage (Fig.~\ref{fig:architecture}J) merges the envelopes of all papers whose extraction succeeded into a single structured database file, together with run identifiers and summary statistics. Per-paper envelope files are retained alongside the merged export, so any database record can be traced back to the individual extraction and, from there, through the manifests, to the source PDF and the original query, all without leaving the run directory.

\subsection{Configuration model and LLM providers}
\label{sec:configuration}

A single YAML file controls the pipeline, with a global LLM default that any stage may override (provider, endpoint, and model). LitCurate supports two API interfaces: the Anthropic API \citep{anthropic2025api} and OpenAI-compatible chat-completions endpoints \citep{openai2025api}. The latter provides access to commercial models, including OpenAI and Gemini models when offered through compatible endpoints, as well as self-hosted open-weight models served using platforms such as Ollama \citep{ollama2024} or vLLM. Providers and models may be combined within a single run, for example, a locally served open-weight model for high-volume screening and a commercial API model for extraction. The configuration, extraction schema, and extraction prompt used for the lower-mantle EOS run are given in Supplementary Text~S1 (Section~S1).

\section{Results}
\label{sec:results}
\subsection{Curation run statistics}
\label{sec:results_funnel}

\begin{table}[htbp]
\centering
\caption{Stage-by-stage outcome of lower-mantle EOS curation run. Stage letters refer to Fig.~\ref{fig:architecture}.}
\label{tab:funnel}
\begin{tabularx}{\textwidth}{l X r}
\toprule
Stage & Outcome & Papers retained \\
\midrule
A\; Query planning & 20 queries generated from the research goal & -- \\
B\; Literature search & 12{,}000 query-result records (20 queries $\times$ 12 year slices $\times$ 50 results) & -- \\
C\; Merge and rank & deduplicated, gated, and ranked candidates & 1{,}846 \\
D\; Abstract screening & 1{,}030 kept (56\%) & 1{,}030 \\
E\; PDF acquisition & HTTP-accessible PDFs retrieved automatically; others ingested manually & 918 \\
F\; Conversion and cleanup & 917 cleaned markdown documents (1 conversion failure) & 917 \\
G\; Pre-extraction gate & 67 papers skipped (no regex signals) & 850 \\
H\; Full-text screening & 312 kept & 312 \\
I\; Structured extraction & 300 validated extractions; 12 failed & 300 \\
J\; Export & 1{,}334 EOS entries from 205 papers (95 empty) & 205 \\
\bottomrule
\end{tabularx}
\end{table}

Table~\ref{tab:funnel} summarizes the lower-mantle EOS curation run from query planning to export. To improve temporal coverage, the 1990--2025 search period was divided into twelve non-overlapping three-year publication windows. Each of the 20 generated search queries was executed independently within every window, with up to 50 results retained for each query--window combination. This produced a maximum of 12{,}000 query-result records before deduplication and ranking ($20$ queries $\times$ $12$ windows $\times$ $50$ results). After merging the search results, screening the candidate papers, acquiring and converting the available PDFs, and performing structured extraction, the compiled database contained 1{,}334 EOS entries from 205 papers. An additional 95 papers produced no exportable records, whereas 12 papers failed during extraction.

\subsection{Evaluation of coverage and extraction}
\label{sec:results_validation}
\label{sec:results_known_papers}

To evaluate end-to-end coverage, we assembled a benchmark of 50 open-access papers known to report EOS parameters for lower-mantle minerals. Of these, 37 were represented in the exported database, corresponding to 74\% benchmark coverage. Among the 13 missing papers, nine were not returned by OpenAlex within the per-query, per-year-window result limit. One appeared in the raw search results but was removed during merge and ranking because OpenAlex provided no abstract. The remaining three reached the screening stages: one was excluded during abstract screening and two during full-text screening. These results indicate that most omissions occurred during literature retrieval rather than extraction.

To quantify extraction quality, a mineral physicist constructed a ground-truth dataset by manually reviewing 50 randomly selected papers from the converted corpus and identifying all in-scope static EOS records (numeric $V_0$, $K_0$, and/or $K_0'$ for the target phases). A second domain expert audited this annotation. The resulting dataset comprised 220 records, each labeled for phase, composition, EOS model, $V_0$, $K_0$, and $K_0'$. Four LLMs then ran the identical extraction stage on these prepared texts, using one shared schema and prompt: three commercial reasoning models (Claude Fable 5, GPT-5.6 Sol High, and Gemini 3.1 Pro Preview) and one open-weight reasoning model (Qwen3.6 27B; \citealp{yang2025qwen3}).

The annotating expert then paired extracted records with this ground truth in a matching interface. Pairing was one-to-one: each ground-truth record was matched to a single extracted record or left unmatched, and unmatched extracted records were counted as extras. Unit differences, reported ranges or uncertainties, and duplicate rows were resolved through expert judgment, without applying a coded numerical tolerance or automated matching rule.

\begin{figure}
    \centering
    \includegraphics[width=\textwidth]{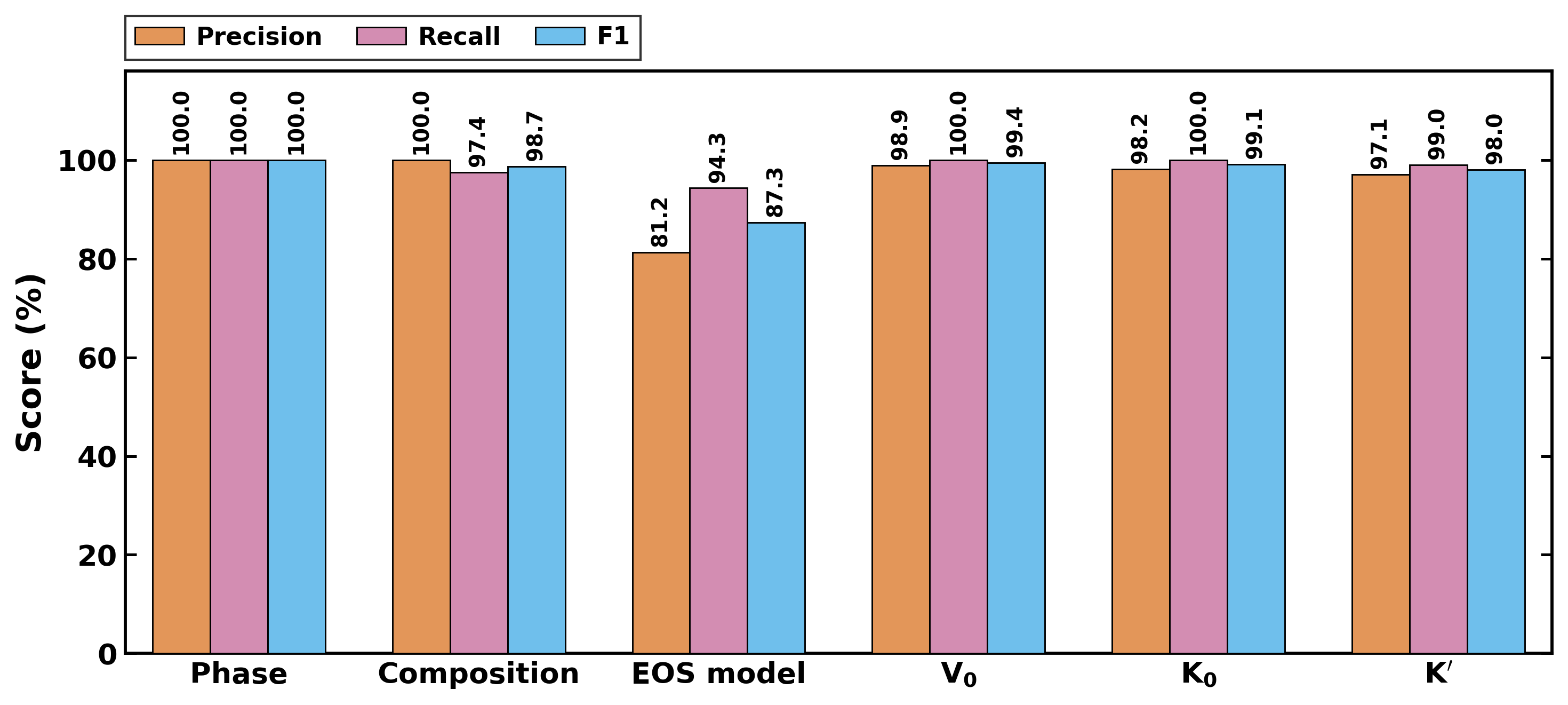}
        \caption{Field-level validation of production extraction model (Claude Fable 5): precision, recall, and F1 per extracted field, computed over matched records.}
    \label{fig:validation_fields}
\end{figure}

Alignment was scored at two levels. At the record level, detection is the fraction of ground-truth records recovered, and the unmatched-record rate is the number of extracted records without a ground-truth counterpart, expressed as a fraction of the ground-truth count. At the field level, each of the six fields on every matched record was scored as a true positive (TP), false positive (FP), true negative (TN), or false negative (FN). A TP was assigned when both the extraction and the ground truth contained a value, and those values agreed (for numerical fields, the reported number; for phase, composition, and EOS model, they matched the annotation). A TN was assigned when both agreed that the field was unreported. An FP was assigned when the model produced a value that the ground truth left empty. An FN was assigned when the ground truth contained a value that the model omitted or that disagreed with the annotation. Precision, recall, F1, and accuracy follow from these counts. Table~\ref{tab:validation_models} reports both levels for each model, together with the mean API cost per paper of extracting these 50 papers, and Fig.~\ref{fig:validation_fields} breaks the field-level performance of the production model down by field.

\begin{table}[htbp]
\centering
\small
\setlength{\tabcolsep}{4pt}
\caption{Evaluation of extraction stage against a manually curated ground truth of 220 EOS records from 50 papers. Detection is share of ground-truth records recovered; unmatched-record rate is share of extracted records without a ground-truth counterpart, relative to ground-truth count. Field-level metrics aggregate six extracted fields over all matched records. Cost is mean API charge (USD) per paper for a single extraction pass over these 50 papers; commercial models were run as reasoning models, which increases token use relative to non-reasoning chat models. Local Qwen3.6 27B deployment incurs no API charge.}
\label{tab:validation_models}
\begin{tabularx}{\textwidth}{>{\raggedright\arraybackslash}X rr rrrr r}
\toprule
 & \multicolumn{2}{c}{Record level (\%)} & \multicolumn{4}{c}{Field level, all fields (\%)} & \\
\cmidrule(lr){2-3}\cmidrule(lr){4-7}
Model & Detection & Unmatched & Accuracy & Precision & Recall & F1 & \shortstack{Cost\\(USD/paper)} \\
\midrule
Claude Fable 5 & 89.1 & 10.0 & 95.7 & 95.9 & 99.4 & 97.6 & $\sim$0.32 \\
Gemini 3.1 Pro Preview & 91.8 & 7.3 & 91.7 & 96.6 & 94.0 & 95.2 & $\sim$0.10 \\
GPT-5.6 Sol High & 95.5 & 23.6 & 95.3 & 96.6 & 98.2 & 97.4 & $\sim$0.21 \\
Qwen3.6 27B & 53.6 & 1.4 & 89.5 & 91.4 & 97.3 & 94.3 & 0 \\
\bottomrule
\end{tabularx}
\end{table}

Claude Fable 5 was used for production extraction. It recovered 89.1\% of the ground-truth records with a 10.0\% unmatched-record rate and achieved the highest field-level accuracy (95.7\%), recall (99.4\%), and F1 score (97.6\%) among the evaluated models. Gemini 3.1 Pro Preview achieved a higher record-detection rate (91.8\%) and a lower unmatched-record rate (7.3\%) at the lowest mean API cost among the commercial models, although its field-level accuracy and F1 score were lower. GPT-5.6 Sol High achieved the highest detection rate (95.5\%) and similarly strong field-level performance, but it also produced the largest number of unmatched records (23.6\%). The open-weight Qwen3.6 27B model incurred no API cost and produced few unmatched records, but its detection rate was substantially lower (53.6\%), indicating that it missed nearly half of the ground-truth records. Its high field-level scores therefore describe only the smaller subset of records that it successfully recovered. Overall, the commercial models provided substantially better record coverage, with different trade-offs among unmatched records, field-level performance, and API cost. These results provide a descriptive comparison of model behavior under the tested configuration rather than a definitive ranking of the models.

Two patterns stand out in the field-level results (Fig.~\ref{fig:validation_fields}). First, the numeric parameters $V_0$, $K_0$, and $K'$ are extracted with high F1 and near-perfect recall, indicating that verbatim transfer of reported values from text and tables is handled reliably. Second, the EOS-model field is the weakest, a pattern shared by the evaluated models, and one that mirrors the reporting gap discussed in Section~\ref{sec:results_database}; papers frequently apply a formulation without naming it alongside the reported values, so the extractor must either recover it from elsewhere in the text or leave it unspecified.

\subsection{The curated database}
\label{sec:results_database}

\begin{figure}
    \centering
    \includegraphics[width=\textwidth]{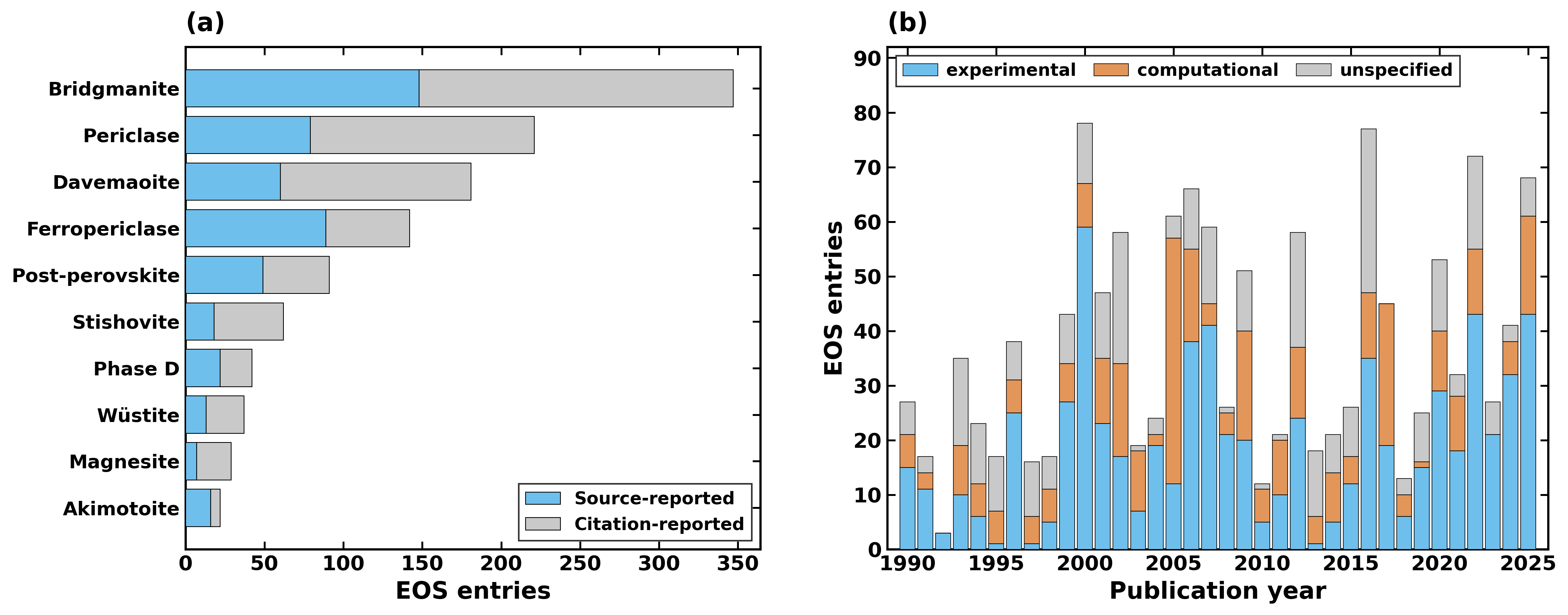}
    \caption{Composition of curated EOS database of lower-mantle and lower-mantle-relevant high-pressure mineral phases (1{,}334 entries from 205 papers). (a)~Entries for ten most frequent phases, divided into source-reported values (new determinations by extracting paper) and citation-reported values (values a paper quotes from earlier literature); four entries of unknown provenance are not shown separately. (b)~Entries per publication year, divided by method class assigned from reported method description (experimental, computational, or unspecified).}
    \label{fig:db_overview}
\end{figure}

The compiled database comprises 1{,}334 EOS entries from 205 papers spanning 1990--2025 (Fig.~\ref{fig:db_overview}). Bridgmanite \citep{tschauner2014discovery} accounts for the largest share (348; 26\%), followed by periclase (221; 16\%), davemaoite (CaSiO$_3$ perovskite; 181; 13\%) \citep{tschauner2021discovery}, ferropericlase (142; 11\%), and post-perovskite (91; 7\%) \citep{murakami2004postperovskite,oganov2004theoretical,tsuchiya2004phase}. Accessory, hydrous, carbon-bearing, and neighboring high-pressure phases (including stishovite, phase~D, magnesite, akimotoite, liebermannite, and calcium-ferrite-type phases) are also present, reflecting recall-oriented retrieval that captured phases reported alongside the primary lower-mantle targets. Classified by their reported method descriptions, 679 entries (51\%) are experimental determinations and 342 (26\%) are computational (DFT and related first-principles methods; \citealp{hohenberg1964inhomogeneous,kohn1965self,karki2000mgo}). The remaining 313 entries (23\%) carry no method because the source paper does not state one: 267 of them are citation-reported values reprinted in compilations or comparison tables; a small remainder comes from thermodynamic-modeling and seismic-inversion papers. Extraction records the method as unspecified when it is not explicitly reported. Field coverage mirrors how EOS parameters are actually reported: $K_0$ is present in 95\% of entries and $K'$ in 82\%, whereas $V_0$ appears in only 59\%, since many studies quote moduli without restating reference volumes.

Beyond the EOS parameters themselves, the database retains the contextual metadata needed for meaningful comparison. Where provenance could be determined, values are classified as source-reported (new determinations made in the paper from which they were extracted; 588 entries) or citation-reported (values reproduced from earlier studies, typically in compilations or comparison tables; 742 entries). The remaining four entries have unknown provenance. Retaining both categories captures how EOS parameters are propagated through the literature while allowing downstream users to restrict analyses to primary determinations when data independence is important (Fig.~\ref{fig:db_overview}a). The database also preserves fitting constraints: $K_0'$ was fitted in 640 entries (48\%), explicitly fixed in 250 (19\%), assumed in 43 (3\%), and not clearly characterized in the remainder. This distinction is essential when comparing $K_0$ values because of the well-known trade-off between $K_0$ and $K_0'$ in EOS fitting \citep{angel2000equations}: a larger fitted $K_0'$ typically goes with a smaller $K_0$, so two studies that both fit $K_0'$ can still report different $K_0$. Where specified, bulk moduli are classified as isothermal or adiabatic (777 and 129 entries, respectively), and the reference temperature and pressure are recorded when reported (607 and 569 entries, respectively). When a paper reports a pressure scale or calibrant, that information is retained in the extra-information object as an optional field, together with other reported quantities such as the Grüneisen parameter ($\gamma_0$), Debye temperature ($\theta_0$), and thermal-expansion coefficient ($\alpha$). Among entries identifying an EOS formulation, third-order Birch--Murnaghan is the most common (434), followed by Birch--Murnaghan fits of unspecified order (163), Vinet (48), and fourth- and second-order Birch--Murnaghan formulations (42 and 33, respectively). Smaller numbers are reported as Murnaghan or generically as finite-strain formulations without further specification. A further 463 entries (35\%) report EOS parameters without identifying the underlying formulation; the database preserves this uncertainty and does not assign a model without supporting evidence. This omission is concentrated among citation-reported values (431 of 463): papers presenting new EOS fits generally identify the formulation used, whereas compilations and textual citations often reproduce parameter values without restating the model of the original fit.

\begin{figure}
    \centering
    \includegraphics[width=\textwidth]{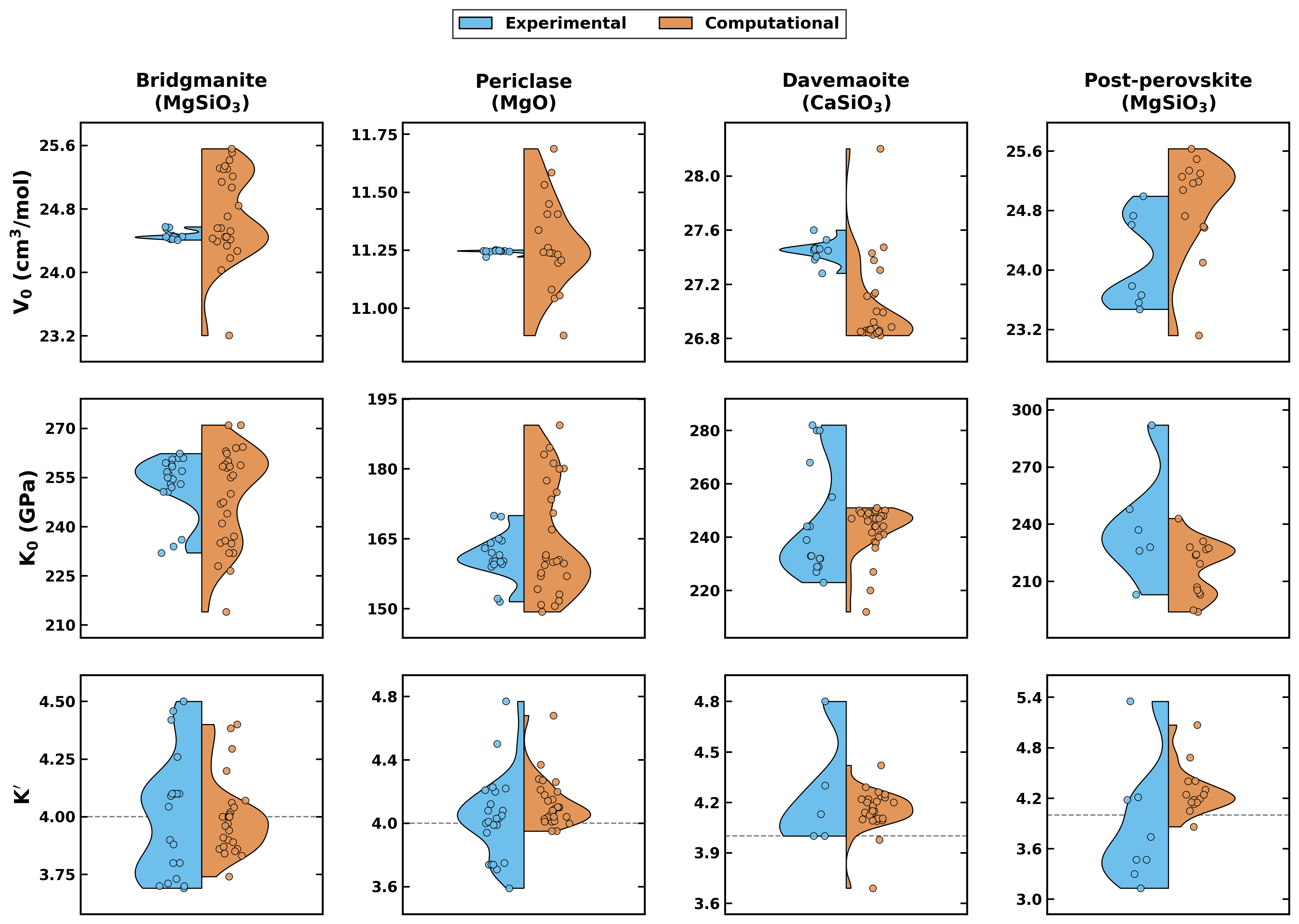}
    \caption{Distributions of extracted EOS parameters for pure endmember bridgmanite (MgSiO$_3$), periclase (MgO), davemaoite (CaSiO$_3$), and post-perovskite (MgSiO$_3$). Solid-solution compositions, citation-reported values, high-temperature fits ($T_{\mathrm{ref}}>310$~K), assumed or fixed parameters, and adiabatic moduli are excluded. Experimental entries (left, blue) are room-temperature static-compression values; computational entries (right, orange) include 0~K static-lattice fits and other calculations with $T_{\mathrm{ref}}\le 310$~K, with no thermal adjustment between these states. Rows are $V_0$ on a common molar-volume basis, $K_0$, and fitted $K'$, with $K'=4$ marked (dashed line).}
    \label{fig:eos_distributions}
\end{figure}

The retained metadata makes it straightforward to isolate comparable subsets of the database, and the resulting distributions provide a qualitative scientific plausibility check on the extracted and normalized values. Fig.~\ref{fig:eos_distributions} shows $V_0$, $K_0$, and $K_0'$ for source-reported static determinations of four frequently reported pure-endmember phases. Volumes are converted to a common molar basis, and experimental and computational values are shown separately. High-temperature fits are excluded. Experimental entries are room-temperature compression measurements; computational entries include both 0~K DFT calculations and calculations reported near room temperature (at or below 310~K). The figure does not convert these to a common temperature, so the two sides should not be read as a matched 0~K versus 300~K comparison. The extracted values reproduce the known behavior of these parameters without any physics-based supervision: for each phase, both methods cluster around consensus values (median $K_0$ of 256 and 255~GPa for experimental and computational bridgmanite, respectively, and 160~GPa for both classes of periclase), and fitted $K'$ concentrates around 4 (median 4.1; 93\% lie between 3.5 and 5). Fixed or assumed $K'$ values are omitted, so the $K'$ row is not dominated by the conventional value of 4. That does not make the remaining $K_0$ values independent of $K'$; the figure shows reported $(K_0,K')$ pairs, not moduli refit to a common $K'$. Differences between the two sides remain visible, such as the wider computational $V_0$ range for post-perovskite, which reflects the choice of DFT functional. Broader computational $K_0$ ranges for davemaoite and post-perovskite likewise include that functional choice and the mixing of 0~K DFT with near-room-temperature calculations.

After export from LitCurate, the extracted records underwent a separate, application-specific post-processing step to standardize heterogeneous terminology, unit labels, and reporting conventions across multiple fields. The $V_0$ fields illustrate this normalization burden: the verbatim extractions contained 33 distinct unit strings, including variant notations for cubic angstroms, molar volumes, cubic nanometres, and atomic units. These labels were mapped onto four standard units (\AA$^3$, cm$^3$\,mol$^{-1}$, nm$^3$, and a.u.$^3$), and each volume was classified by its reported basis as a unit-cell, per-formula-unit, molar, or unknown-basis volume. Comparable normalization was applied to other fields where variant terminology or formatting referred to the same underlying concept. This post-processing was performed outside the LitCurate pipeline; the extracted numeric values and verbatim evidence excerpts were retained, allowing each reported value to be traced to the source text.

\section{Discussion}
\label{sec:discussion}

LitCurate is a reusable, configuration-driven curation package. Domain knowledge (research goals, screening criteria, schemas, and prompts) lives in the configuration, so this discovery-to-export architecture is designed to be adapted to new literature-derived database tasks through configuration rather than modification of the core stage logic. Existing platforms already accelerate extraction once a document collection is in hand \citep{peters2017geodeepdive,zhang2023geodeepshovel,guo2024geoknowledgefusion,wang2024scidasynth,li2025sciex}. LitCurate occupies the complementary gap: it starts from a research goal, screens at two text granularities before expensive extraction, records every keep-or-drop decision, and exports a schema-validated database with per-record provenance.

The lower-mantle application demonstrates how the stage-wise architecture concentrates expensive LLM extraction on the most relevant documents. Most of the 12{,}000 OpenAlex query-result records did not reach the extraction stage because merge-and-rank, abstract screening, the pre-extraction signal gate, and full-text screening progressively reduced the candidate set. That funnel also explains missed papers: on the 50-paper open-access benchmark (Section~\ref{sec:results_known_papers}), 37 studies appear in the exported database, and most of the remainder were missed at retrieval, so a relevant paper can be absent even when extraction on converted text is accurate. A locally served open-weight model can do the high-volume screening at no API cost, while a commercial model is reserved for the papers that survive screening; Table~\ref{tab:validation_models} shows that several commercial models perform that extraction stage well enough for production use.

EOS parameters set the reference state for mineral compression and are used to compare experimental and first-principles results and to calculate densities at lower-mantle pressures. The retained phase, composition, method, formulation, unit, and fitting-constraint metadata substantially reduce the effort required to identify and normalize comparable determinations. Source-reported and citation-reported labels allow users to distinguish primary determinations from values reproduced from earlier studies and, when data independence is required, to restrict analyses to source-reported records or apply further deduplication. The database identifies phases and compositions that are sparsely represented within the curated corpus. These cases can guide targeted literature audits and, once coverage has been verified, may help identify priorities for future experiments and calculations. The records can be explored through a searchable web application.

Three limits follow from this design. Extraction reads converted text and tables, so values that appear only as plotted symbols, unlabeled curves, or other figure content are missed; automated figure extraction is not yet mature enough for scientific database construction \citep{he2026exchart,davila2021chart,masry2022chartqa,wang2024charxiv}, but can be added later as a stage without rewriting discovery, screening, or export. Automatic PDF download is limited to open-access locations returned by Unpaywall and OpenAlex, and an open-access flag does not always resolve to a downloadable file, so copies missed by automated retrieval can be added manually. Post-processing is left outside the pipeline because unit conversion and related normalization are domain-specific. Encoding those rules in LitCurate would hard-code one discipline's conventions into a package intended to be reused through configuration alone. Records are therefore exported as reported, allowing users to apply the conversions required for their analyses.

\section{Conclusions}
\label{sec:conclusions}

We have presented LitCurate, an open-source framework that transforms a research objective into a structured, schema-validated literature database through a transparent sequence of inspectable stages. Domain-specific knowledge is defined through configurable research goals, screening criteria, schemas, and prompts, allowing the workflow to be adapted to different scientific questions without modifying the core pipeline. Support for multiple LLM providers, persistent intermediate outputs, and a detailed run ledger makes large curation runs resumable, traceable, and auditable.

Applied to EOS parameters for lower-mantle minerals and related high-pressure phases, LitCurate produced 1{,}334 entries from 205 papers. Each entry links one or more reported EOS parameters ($V_0$, $K_0$, and/or $K_0'$) to the corresponding phase, composition, method, EOS formulation, fitting constraints, and literature provenance, including whether the value is source-reported, citation-reported, or of unknown provenance. End-to-end evaluation against 50 open-access papers known to contain relevant EOS parameters found that 37 were represented in the exported database, corresponding to 74\% coverage, with most omissions occurring during literature retrieval. On a separate, domain-expert-curated ground-truth dataset of 50 randomly selected papers, the production extractor recovered 89.1\% of EOS records with a 10.0\% unmatched-record rate. Across matched records, it achieved 95.7\% field-level accuracy and an F1 score of 97.6\%.

The present EOS application is an initial demonstration of the framework. Future schemas could extend extraction to elastic tensors, polycrystalline moduli, seismic velocities, thermal and thermoelastic properties, phase boundaries, compositional partitioning, and transport properties. Another important extension is quantitative extraction from figures. Because current models are not yet sufficiently accurate for complex scientific plots, this capability remains outside the present workflow. Adding this capability as models improve, together with uncertainty estimates and expert validation, could substantially increase the coverage and scientific utility of LitCurate.

The resulting records are available as a deposited database and through a searchable web application, providing a traceable resource for comparing EOS determinations across phases, compositions, methods, and fitting choices. More broadly, LitCurate keeps the core pipeline reusable by expressing domain-specific requirements through configuration. Adapting the framework to a new literature-curation target primarily requires defining the relevant research goals, screening criteria, extraction schemas, and prompts while retaining the core discovery-to-export architecture.

\section*{Declaration of competing interests}

The authors declare no competing interests.

\section*{Funding}
This work was supported primarily by the Gordon and Betty Moore Foundation (Award GBMF12801; \url{https://doi.org/10.37807/GBMF12801}). RMW and WS also received support from the U.S. National Science Foundation (Award EAR-2506448).

\section*{Code availability}

{\parindent=0pt
Name of the code/library: LitCurate

Contact: Abin Shakya (ashakya@ldeo.columbia.edu)

Hardware requirements: CPU; CUDA-capable GPU recommended for large-scale PDF conversion

Program language: Python 3.10 or later

Software required: Python package dependencies listed in the repository; an Anthropic or OpenAI-compatible LLM API (optional local serving via Ollama or vLLM); Marker for PDF-to-markdown conversion

Program size: see repository

Source code is available at \url{https://github.com/MineralsCloud/litcurate} under the MIT License.

Documentation: \url{https://litcurate.com}
} 

\section*{Data availability}

The lower-mantle equation-of-state database compiled in this study (1{,}334 entries from 205 papers) is available at Zenodo: \url{https://doi.org/10.5281/zenodo.22118629}. Column definitions for the deposited data are provided in Table~\ref{tab:s_db_fields}. The database can also be browsed online at \url{https://eos.litcurate.com}.

\section*{Author contributions}
Abin Shakya: Conceptualization, Methodology, Software, Data curation, Formal analysis, Writing -- original draft.
Wilson Samuels: Software, Data curation, Formal analysis, Writing -- review and editing.
Dominica Wilson: Software, Data curation, Formal analysis, Writing -- review and editing.
Gioia A. Marchi: Software, Data curation, Formal analysis, Writing -- review and editing.
Israa Draz: Data curation, Validation, Writing -- review and editing.
Chenxing Luo: Conceptualization, Investigation, Validation, Writing -- review and editing.
Renata M. Wentzcovitch: Funding acquisition, Conceptualization, Investigation, Writing -- review and editing.

\bibliographystyle{apalike}
\bibliography{bibliography}

\clearpage
\setcounter{secnumdepth}{0}
\renewcommand{\thefigure}{S\arabic{figure}}
\renewcommand{\thetable}{S\arabic{table}}
\setcounter{figure}{0}
\setcounter{table}{0}

\section*{Supplementary material}

\begin{figure}[H]
    \centering
    \includegraphics[
        width=\textwidth,
        height=0.78\textheight,
        keepaspectratio
    ]{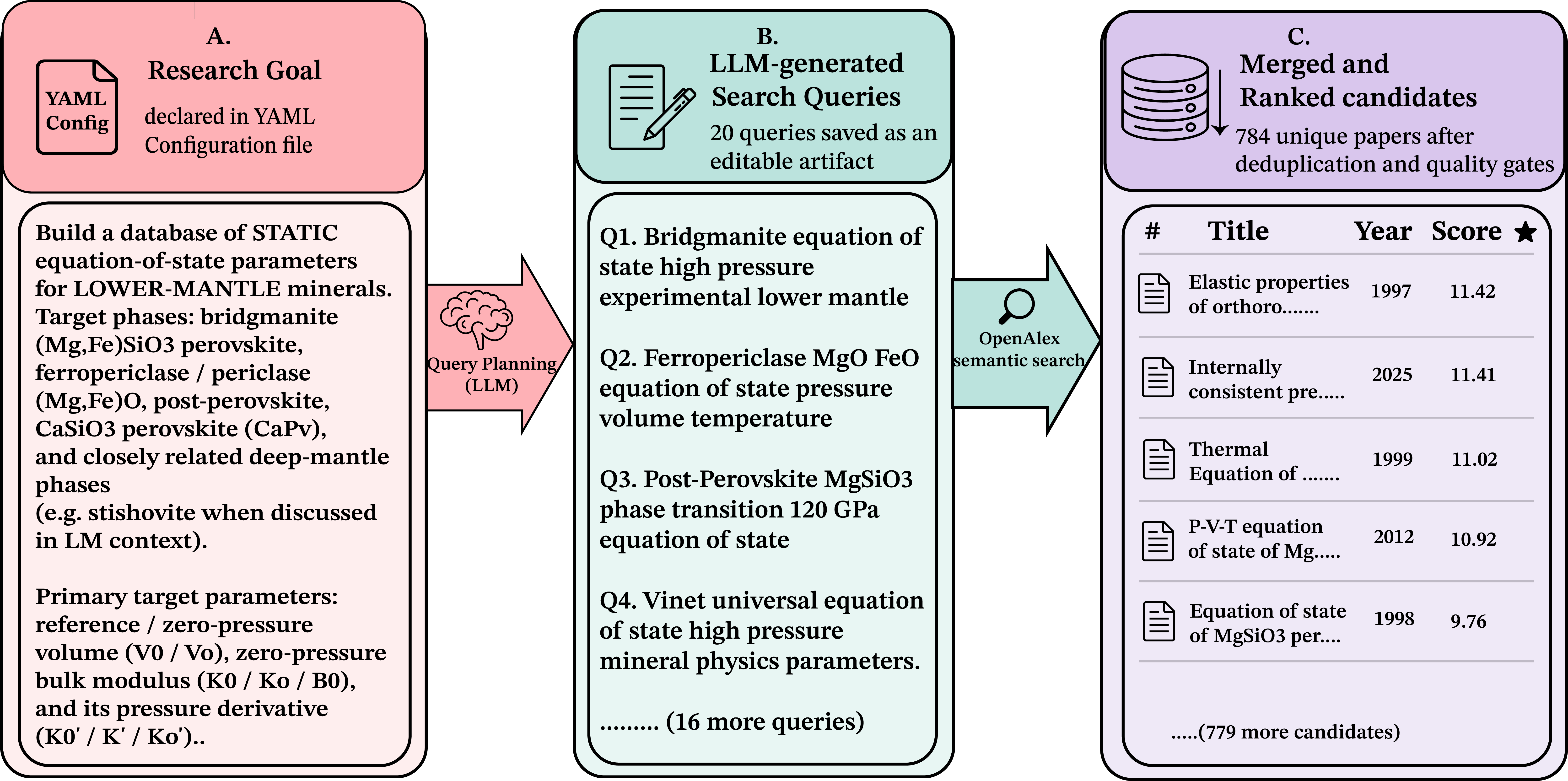}
    \caption{Detailed view of literature discovery (stages A--C of main-text architecture figure). Research goal is expanded into a portfolio of search queries (A), each query is executed against OpenAlex over publication-year slices (B), and overlapping results are merged, quality-gated, and ranked into a single table of candidate papers (C).}
    \label{fig:s_discovery}
\end{figure}

\begin{figure}[H]
    \centering
    \includegraphics[
        width=\textwidth,
        height=0.78\textheight,
        keepaspectratio
    ]{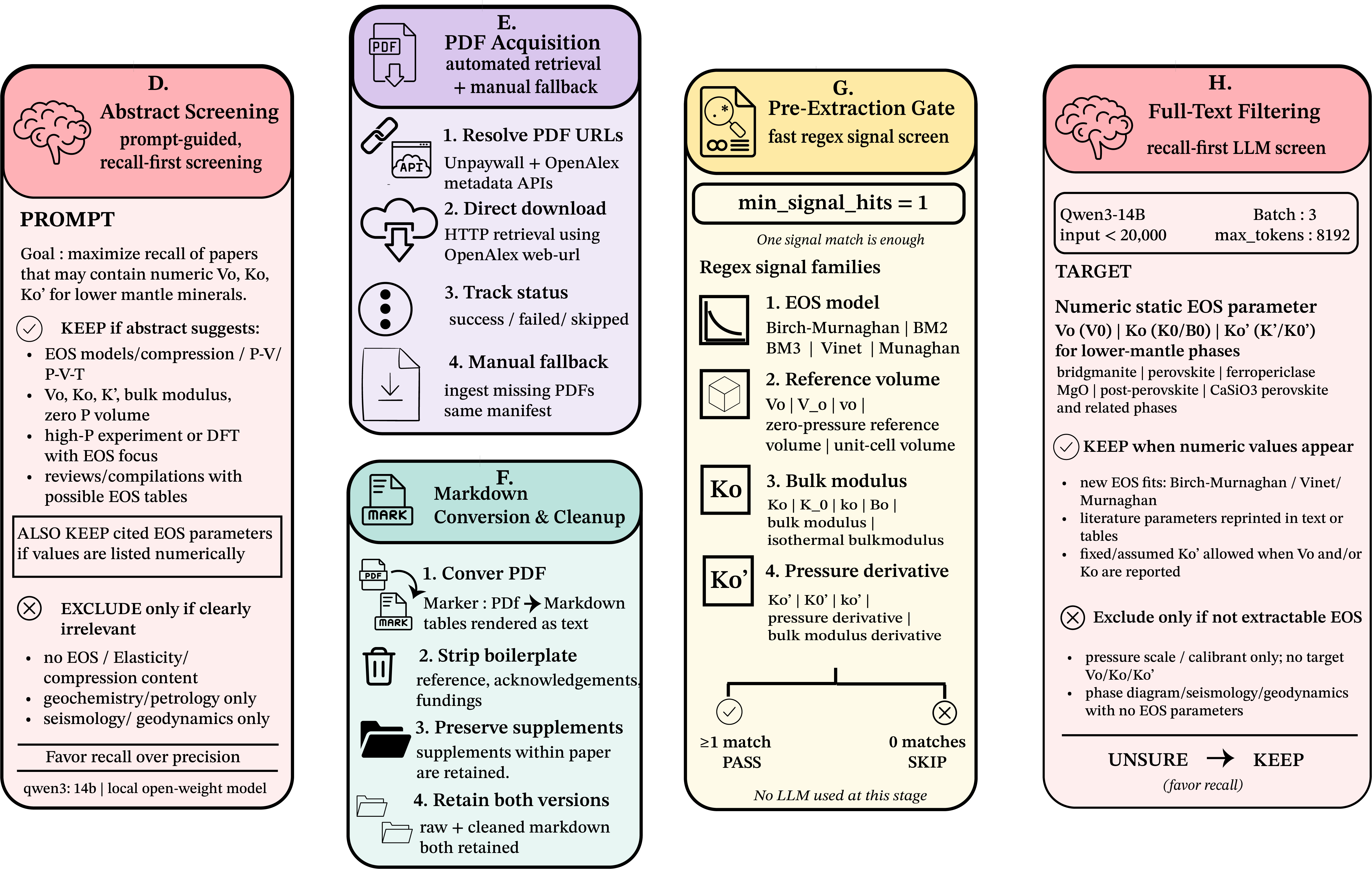}
    \caption{Detailed view of screening and full-text preparation (stages D--H of main-text architecture figure). Candidates are screened at abstract level (D), PDFs are acquired (E) and converted to cleaned markdown (F), a regular-expression pre-extraction gate skips documents with no EOS-parameter signals (G), and remaining papers are screened at full-text level (H).}
    \label{fig:s_screening}
\end{figure}

\section*{S1. Configuration, schema, and extraction prompt}
\label{sec:si_config}

The files below are those used to compile the lower-mantle static EOS database. Listing~S1 is the run configuration (\path{configs/minerals_eos.yaml}); Listing~S2 is the extraction schema (\path{schemas/minerals/eos_reported.yaml}); Listing~S3 is the extraction prompt (\path{prompts/minerals/eos_reported_static_lm.md}). A new literature-derived database is obtained by replacing these three files rather than rewriting pipeline stages.

\singlespacing
\paragraph{Listing S1.} Run configuration used in this study (\path{configs/minerals_eos.yaml}).
\begin{lstlisting}
run:
  name: lower_mantle_static_eos_v2
  user_goal: |
    Build a database of STATIC equation-of-state parameters for LOWER-MANTLE minerals.
    Target phases: bridgmanite (Mg,Fe)SiO3 perovskite, ferropericlase / periclase
    (Mg,Fe)O, post-perovskite, CaSiO3 perovskite (CaPv), and closely related deep-mantle
    phases (e.g. stishovite when discussed in LM context).

    Primary target parameters: reference / zero-pressure volume (V0 / Vo), zero-pressure
    bulk modulus (K0 / Ko / B0), and its pressure derivative (K0' / K' / Ko').
    Include both newly fitted values AND values reported from other studies when the
    paper lists numeric V0, K0, and/or K0' in text or tables (with provenance when given).

llm:
  provider: anthropic

search:
  num_queries: 20
  provider: anthropic
  model: claude-sonnet-4-6
  retrieval_objective: |
    Retrieve papers likely to report NUMERIC static EOS parameters V0, K0, and/or K0'
    (aliases: Vo, Ko, B0, K', K'0, etc.) for lower-mantle phases: bridgmanite /
    MgSiO3 or (Mg,Fe)SiO3 perovskite, ferropericlase / magnesiowüstite / periclase /
    (Mg,Fe)O / MgO, post-perovskite (PPv), and CaSiO3 perovskite.

    Prioritize Birch-Murnaghan, Vinet, or Murnaghan (and equivalent) static / isothermal
    EOS fits and tables of EOS parameters. INCLUDE:
    - primary experimental or theoretical fits from this study
    - compilations, reviews, and modeling papers that TABULATE or quote V0, K0, K0'
      values for LM minerals (even if taken from earlier literature)
    - papers with P-V or P-V-T data reduced to static EOS parameters

    Deprioritize (but do not exclusively optimize against) purely seismic inversions,
    geodynamic applications with no parameter tables, and pressure-scale-only uses of EOS.
  merge:
    strategy: union_dedupe_rank
    min_query_hits: 1
    top_k: 2000
    frequency_weight: 1.0
    relevance_weight: 1.0
    citation_weight: 0.1
  openalex:
    max_results_per_query: 50
    per_page: 50
    year_slices:
      - year_min: 1990
        year_max: 1992
      - year_min: 1993
        year_max: 1995
      - year_min: 1996
        year_max: 1998
      - year_min: 1999
        year_max: 2001
      - year_min: 2002
        year_max: 2004
      - year_min: 2005
        year_max: 2007
      - year_min: 2008
        year_max: 2010
      - year_min: 2011
        year_max: 2013
      - year_min: 2014
        year_max: 2016
      - year_min: 2017
        year_max: 2019
      - year_min: 2020
        year_max: 2022
      - year_min: 2023
        year_max: 2025
    request_delay_seconds: 1.0

abstract_filter:
  criteria: |
    Goal: maximize recall of papers that may contain NUMERIC static EOS parameters
    V0, K0, and/or K0' (aliases Vo, Ko, B0, K', K'0, …) for LOWER-MANTLE minerals
    (bridgmanite / Mg-silicate perovskite, ferropericlase/periclase/(Mg,Fe)O, 
    post-perovskite, CaSiO3 perovskite, related deep-mantle phases).

    KEEP if the abstract suggests any of:
    - EOS, Birch-Murnaghan, Vinet, Murnaghan, compression, P-V or P-V-T fitting
    - bulk modulus, K0, K', V0, zero-pressure volume for these phases
    - high-pressure experiments or DFT/ab initio on these LM phases with elastic/EOS focus
    - reviews or compilations that may tabulate EOS parameters for mantle minerals

    Numeric values need NOT appear in the abstract.

    ALSO KEEP papers that use or cite previously published EOS parameters if they may
    still list those parameters numerically in tables or text (compilations, comparisons,
    thermoelastic models). Do NOT drop solely because parameters may come from earlier work.

    EXCLUDE only when clearly irrelevant:
    - no connection to mineral EOS / elasticity / compression of LM or high-P mantle phases
    - pure geochemistry/petrology with no elastic/EOS content
    - pure seismology/geodynamics with no mineral parameter tables likely

    Favor recall over precision.

  provider: openai_compatible
  base_url: http://localhost:11434/v1
  model: qwen3:14b
  batch_size: 5
  max_tokens: 8192

download:
  # Attempt metadata-provided PDF URLs over HTTP (not OA-flag-only).
  # Paywalled PDFs will still often fail — use litcurate ingest-pdfs for those.
  unpaywalled_only: false
  request_delay_seconds: 1.0

conversion:
  engine: marker
  device: cuda:0

markdown_clean:
  strip_references: true
  strip_acknowledgments: true
  # EOS tables are frequently in supplements — keep them for extraction.
  strip_supplementary: false
  strip_funding: true

pre_extract:
  enabled: true
  # One strong signal is enough (OCR often drops a second alias).
  min_signal_hits: 1
  regex_signals:
    - "(?i)Birch[- ]?Murnaghan|BM2|BM3|Vinet|Murnaghan|equation of state|\\bEOS\\b"
    - "(?i)\\bV\\s*0\\b|\\bV_0\\b|\\bVo\\b|zero[- ]pressure volume|reference volume|unit[- ]cell volume"
    - "(?i)\\bK\\s*0\\b|\\bK_0\\b|\\bKo\\b|\\bB\\s*0\\b|bulk modulus|isothermal bulk modulus"
    - "(?i)K['′]?\\s*0|K_0['′]?|\\bK['′]\\b|\\bKo['′]?\\b|pressure derivative|bulk modulus derivative"

fulltext_filter:
  enabled: true
  provider: openai_compatible
  base_url: http://localhost:11434/v1
  model: qwen3:14b
  batch_size: 3
  max_tokens: 8192
  # Look further into the paper; tables of V0/K0/K' are often not in the first pages.
  input_max_chars: 20000
  criteria: |
    KEEP if the excerpt contains (or strongly indicates) NUMERIC static EOS parameters
    useful for a lower-mantle database, especially V0 (Vo), K0 (Ko/B0), and/or K0' (K'/Ko'),
    for bridgmanite / (Mg,Fe)SiO3 perovskite, ferropericlase / (Mg,Fe)O / MgO / periclase,
    post-perovskite, CaSiO3 perovskite, or closely related deep-mantle phases.

    KEEP all of the following when numeric parameters appear or are clearly tabulated:
    - New fits from THIS study (Birch-Murnaghan, Vinet, Murnaghan, etc.)
    - Parameters taken FROM OTHER STUDIES if the paper lists numeric V0, K0, and/or K0'
      in text or tables (compilations, comparisons, reviews with parameter tables,
      models that reprint literature EOS constants)
    - Cases where K0' is fixed/assumed (e.g. K'=4) while V0 and/or K0 are reported

    EXCLUDE only if there is no plausible extractable static EOS parameter content, e.g.:
    - EOS mentioned only as a pressure scale / calibrant with no LM mineral V0/K0/K' values
    - Ambient moduli only, with no high-P EOS parameters
    - Phase diagram / seismology / geodynamics with no mineral EOS numbers

    When unsure but V0/K0/K'-like numbers for LM phases might exist, KEEP (favor recall).

extraction:
  # Use static-LM prompt that KEEPS literature-tabulated V0/K0/K' as well as new fits.
  provider: anthropic
  model: claude-fable-5
  section_mode: cleaned_markdown
  markdown_max_chars: 80000
  schemas:
    - name: eos_reported
      format: declarative_yaml
      path: schemas/minerals/eos_reported.yaml
      prompt: prompts/minerals/eos_reported_static_lm.md
      version: "2.1"
      empty_list_field: eos_entries
    - name: source
      path: schemas/minerals/source.json
      fill_from: papers_meta
      skip_when_prior_empty: eos_reported

dry_run: false
\end{lstlisting}

\paragraph{Listing S2.} Extraction schema used in this study (\path{schemas/minerals/eos_reported.yaml}).
\begin{lstlisting}
title: EOSReportedList
version: "2.1"
$id: litcurate://schemas/minerals/eos_reported/2.1
description: >
  Flat reported-layer EOS extraction focused on static V0, K0, and Kp (K').
  Values and units are verbatim from the paper. Canonicalization is downstream.

fields:
  eos_entries:
    array:
      required: [phase, eos_model, evidence]
      fields:
        phase: string
        composition: string?
        structure: string?
        sample: string?

        eos_model: string
        method: string?

        V0: string|number?
        V0_unit: string?
        V0_name: string?
        V0_basis: enum(unit_cell, molar, unknown)?
        V0_determination: enum(measured, fitted, assumed, unknown)?

        K0: string|number?
        K0_unit: string?
        K0_name: string?
        K0_type: enum(isothermal, adiabatic, unknown)?
        K0_determination: enum(fitted, measured, assumed, unknown)?

        Kp: string|number?
        Kp_unit: string?
        Kp_name: string?
        Kp_determination: enum(fitted, fixed, assumed, unknown)?

        T_ref: string|number?
        T_ref_unit: string?
        P_ref: string|number?
        P_ref_unit: string?

        origin: enum(this_study, cited, unknown)?

        evidence: string
        confidence:
          type: number
          minimum: 0
          maximum: 1
          optional: true

        extra_info:
          type: object
          additionalProperties: true
          optional: true
\end{lstlisting}

\paragraph{Listing S3.} Extraction prompt used in this study (\path{prompts/minerals/eos_reported_static_lm.md}).
\begin{lstlisting}
Extract STATIC equation-of-state (EOS) parameters for lower-mantle-relevant phases
into the flat schema. Priority fields: **V0, K0, and Kp (K′)**. Values and units
must be copied exactly as reported.

Target phases (keep these; skip unrelated crustal/transition-zone-only phases unless
explicitly framed as lower-mantle):
- bridgmanite / (Mg,Fe)SiO3 or MgSiO3 perovskite (Pv)
- ferropericlase / magnesiowüstite / periclase / (Mg,Fe)O / MgO
- post-perovskite (PPv)
- CaSiO3 perovskite (CaPv)
- closely related deep-mantle SiO2 or hydrous phases only when discussed with EOS numbers

For each distinct phase + composition + sample/table row + EOS model, return one entry.

Promote when present:
- `V0` (+ `V0_unit`, `V0_name`, `V0_basis`, `V0_determination`)
- `K0` (+ `K0_unit`, `K0_name`, `K0_type`, `K0_determination`)
- `Kp` for K′ / K0′ / K_T′ (+ `Kp_name`, `Kp_determination`)

Determination enums:
- `V0_determination`: `measured` | `fitted` | `assumed` | `unknown`
- `K0_determination`: `fitted` | `measured` | `assumed` | `unknown`
- `Kp_determination`: `fitted` | `fixed` | `assumed` | `unknown`

Ambient / Table zero-pressure unit-cell volumes used with an EOS fit → fill `V0`
with `V0_determination: measured` even if the text only quotes fitted K0 and fixed
K′. Do not park those volumes only in `extra_info`.

Also set: `eos_model`, `method`, `T_ref`/`P_ref` (+ units), `origin`
(`this_study` | `cited` | `unknown`).

INCLUDE literature-tabulated V0/K0/K′ when reported numerically (`origin: cited`).
Other parameters (γ0, θ0, q, α, …) → `extra_info` only.
`evidence` is one string. Do not invent values or convert units.
If no relevant static EOS parameters appear, return `{ "eos_entries": [] }`.

Return JSON matching the schema: `{ "eos_entries": [ ... ] }`.
\end{lstlisting}
\onehalfspacing

\section*{S2. Zenodo xlsx deposit}

Table~S1 defines the columns of the Excel file deposited on Zenodo (1{,}334 entries from 205 papers, 1990--2025). Values in columns named \texttt{Reported*} are stored as extracted from the paper. \texttt{V0(cm3/mol)} and \texttt{K0(GPa)} are post-export conversions from the declared unit only. Empty phase names are recorded as Unspecified.

\singlespacing
\setlength{\LTcapwidth}{\textwidth}
\begin{longtable}{>{\raggedright\arraybackslash}p{0.30\textwidth}>{\raggedright\arraybackslash}p{0.64\textwidth}}
\caption{Column dictionary for Zenodo xlsx deposit.}
\label{tab:s_db_fields}\\
\hline
Field & Description \\
\hline
\endfirsthead
\caption[]{Column dictionary for Zenodo xlsx deposit (continued).}\\
\hline
Field & Description \\
\hline
\endhead
\hline
\endfoot
\texttt{doi} & Digital Object Identifier of the source paper. \\
\texttt{title} & Paper title. \\
\texttt{year} & Publication year. \\
\texttt{phase} & Mineral or high-pressure phase name. Unspecified if the paper does not name a phase. \\
\texttt{composition} & Chemical composition as reported. \\
\texttt{structure} & Crystal structure, when stated. \\
\texttt{sample} & Sample or specimen description, when stated. \\
\texttt{eos\_model} & EOS formulation named in the paper (for example third-order Birch--Murnaghan), or unknown. \\
\texttt{method} & Experimental, Computational, or Unspecified, assigned from the reported method description. \\
\texttt{method\_reported} & Original method string from extraction. \\
\texttt{origin} & Source-reported (new determination in the extracting paper), citation-reported (value quoted from earlier literature), or Unspecified. \\
\texttt{Reported V0} & Reference volume as reported. \\
\texttt{Reported V0\_unit} & Unit given for the reported $V_0$ (for example \AA$^3$ or cm$^3$/mol). \\
\texttt{Reported V0\_basis} & Volume basis: unit cell, molar, or unknown. \\
\texttt{Reported V0\_determination} & How $V_0$ was obtained (fitted, assumed, and related labels). \\
\texttt{V0(cm3/mol)} & $V_0$ converted to cm$^3$/mol from the declared unit and basis, when conversion is possible. \\
\texttt{Reported K0} & Bulk modulus as reported. \\
\texttt{Reported K0\_unit} & Unit given for the reported $K_0$ (GPa, kbar, Mbar, or TPa). \\
\texttt{Reported K0\_type} & Isothermal, adiabatic, or unknown. \\
\texttt{Reported K0\_determination} & How $K_0$ was obtained (fitted, measured, assumed, or unknown). \\
\texttt{K0(GPa)} & $K_0$ converted to GPa from the declared unit (1\,TPa $=$ 1000\,GPa; 1\,Mbar $=$ 100\,GPa; 1\,kbar $=$ 0.1\,GPa). \\
\texttt{Reported Kp} & Pressure derivative $K'$ as reported. \\
\texttt{Reported Kp\_determination} & Fitted, fixed, assumed, or unknown. \\
\texttt{T\_ref} & Reference temperature, when reported. \\
\texttt{T\_ref\_unit} & Unit of \texttt{T\_ref}. \\
\texttt{P\_ref} & Reference pressure, when reported. \\
\texttt{P\_ref\_unit} & Unit of \texttt{P\_ref}. \\
\texttt{evidence} & Quoted text or table excerpt supporting the extracted values. \\
\texttt{extra\_info} & Optional additional quantities reported with the EOS (JSON object). \\
\end{longtable}
\doublespacing

\end{document}